\documentclass[twocolumn,showpacs,preprintnumbers,amsmath,amssymb]{revtex4}

\usepackage{graphicx}
\usepackage{epsfig}
\usepackage{dcolumn}
\usepackage{bm}
\usepackage{tabularx}

\newcommand{\bei}{\begin{itemize}}
\newcommand{\eei}{\end{itemize}}
\newcommand{\beq}{\begin{equation}}
\newcommand{\eeq}{\end{equation}}
\newcommand{\beqn}{\begin{eqnarray}}
\newcommand{\eeqn}{\end{eqnarray}}
\newcommand{\beqns}{\begin{eqnarray*}}
\newcommand{\eeqns}{\end{eqnarray*}}

\newcommand{\equaref}[1]{Eq.~(\ref{eq:#1})}

\newcommand{\figref}[1]{Fig.~\ref{fig:#1}}

\newcommand{\mueV} {\mu\rm{eV}}
\newcommand{\meV} {\rm{meV}}
\newcommand{\MeV} {\rm{MeV}}
\newcommand{\GeV} {\rm{GeV}}

\newcommand{\Mpara} {m_{\gamma'}}

\newcommand{\Pem} {P_{\rm{em}}}
\newcommand{\Pdet} {P_{\rm{det}}}

\newcommand{\TMmode} {$\rm{TM_{010}}$}

\def\ea{{\em et al.}}

\newcommand{\jprlBase}       {Phys.\ Rev.\ Lett.}
\newcommand{\jprBase}        {Phys.\ Rev.}
\newcommand{\jplBase}        {Phys.\ Lett.}
\newcommand{\nimBaseA}       {Nucl.\ Instrum.\ Methods Phys.\ Res., Sect.\ A}

\newcommand{\aplet}     [1]  {{Appl.\ Phys.\ Lett.\ {\bf #1}}}

\newcommand{\jetp}      [1]  {{JETP~{\bf #1}}}

\newcommand{\jap}       [1]  {{J.\ Appl.\ Phys.\ {\bf #1}}}

\newcommand{\nature}       [1]  {Nature~{\bf #1}}
\newcommand{\nima}      [1]  {\nimBaseA~{\bf #1}}

\newcommand{\plb}       [1]  {\jplBase\ B~{\bf #1}}

\newcommand{\jprl}      [1]  {\jprlBase\ {\bf #1}}
\newcommand{\jprd}      [1]  {\jprBase\ D~{\bf #1}}

\newcommand{\rsi}       [1]  {{Rev.\ Sci.\ Instr.\ {\bf #1}}}

\newcommand{\spjetp}    [1]  {{Sov.\ Phys.\ \jetp{#1}}}

\begin{document}

\title{A Search for Hidden Sector Photons with ADMX}

\author{A. Wagner, G. Rybka, M. Hotz, L. J Rosenberg}
\affiliation{University of Washington, Seattle, Washington 98195}
\author{S.J. Asztalos\footnote{Currently at XIA LLC, 31057 Genstar Rd., Hayward CA, 94544.}, G. Carosi, C. Hagmann, D. Kinion and K. van Bibber\footnote{Currently at Naval Postgraduate School, Monterey CA, 93943.}}
\affiliation{Lawrence Livermore National Laboratory, Livermore, California, 94550}
\author{J. Hoskins, C. Martin, P. Sikivie and D.B. Tanner}
\affiliation{University of Florida, Gainesville, Florida 32611}
\author{R. Bradley}
\affiliation{National Radio Astronomy Observatory, Charlottesville, Virginia 22903}
\author{J. Clarke}
\affiliation{University of California and Lawrence Berkeley National Laboratory, Berkeley, California 94720}

\date{\today}

\begin{abstract}
Hidden U(1) gauge symmetries are common to many extensions of the Standard Model proposed to explain dark matter. The hidden gauge vector bosons of such extensions may mix kinetically with Standard Model photons, providing a means for electromagnetic power to pass through conducting barriers. The ADMX detector was used to search for hidden vector bosons originating in an emitter cavity driven with microwave power. We exclude hidden vector bosons with kinetic couplings $\chi>3.48\times 10^{-8}$ for masses less than $3~\mueV$. This limit represents an improvement of more than two orders of magnitude in sensitivity relative to previous cavity experiments.  
\end{abstract}


\maketitle
The Standard Model (SM) of particle physics explains the electromagnetic, weak and strong interactions of the leptons and quarks that comprise the visible universe. Extensions of the SM allow for a so-called hidden sector of particles; these particles may be so extremely weakly interacting with the known SM particles that they have little effect on observations. This hidden sector has been suggested as a model for dark matter in the universe~\cite{HamedWeiner} where the particles of the hidden sector are charged under a hidden U(1) gauge symmetry. The hidden U(1) vector boson may mix kinetically~\cite{Holdom} with the SM photon providing a weak coupling of the hidden sector to the SM. In those models where the kinetic mixing provides the only coupling of the hidden sector to the SM, the hidden vector boson is referred to as a paraphoton. Paraphotons with a mass between $50~\MeV$ and $1~\GeV$ could provide a mechanism to explain the excess positron fraction observed by PAMELA~\cite{PAMELA} while light paraphotons in the mass range $1~\mueV - 1~\meV$ have been shown to produce a hidden cosmic microwave background (hCMB)~\cite{hCMB}. 

Paraphoton-photon mixing may be detected directly via the measurement of transmitted power between shielded microwave cavities~\cite{JackelRing}. We use the Axion Dark Matter eXperiment (ADMX) detector in this way to search for evidence of hidden sector kinetic mixing. This experiment is sensitive to paraphotons with $\Mpara\approx 3\mueV$, the low end of the paraphoton mass range that may produce a hCMB.  

In general the Lagrangian describing the SM and a hidden sector U(1) with kinetic coupling $\chi$ is~\cite{JackelRing} 

\begin{multline}
\label{eq:L}
{\cal L} = -\frac{1}{4}F^{\mu\nu}F_{\mu\nu} -{1\over 4}B^{\mu\nu}B_{\mu\nu} \\
-{1\over 2}\chi F^{\mu\nu}B_{\mu\nu} +{1\over 2}\Mpara^2 B^{\mu}B_{\mu}, 
\end{multline}

\noindent where $B^\mu$ is the vector potential of the paraphoton, $\Mpara$ is the mass of the paraphoton, and $F^{\mu\nu}$ and $B^{\mu\nu}$ are the electromagnetic and hidden sector field strength tensors, respectively. The kinetic coupling described by the $F^{\mu\nu}B_{\mu\nu}$ term in~\equaref{L} results in a mixing of photons with paraphotons. The origin of this mixing is made explicit with the change of basis $B^\mu\rightarrow \tilde{B}^\mu - \chi A^\mu$ to diagonalize the kinetic terms in~\equaref{L}, 

\begin{multline}
\label{eq:Lint}
{\cal L} = -\frac{1}{4}F^{\mu\nu}F_{\mu\nu} -{1\over 4}\tilde{B}^{\mu\nu}\tilde{B}_{\mu\nu} \\
+{1\over 2}\Mpara^2(\tilde{B}^{\mu}\tilde{B}_{\mu} - 2\chi\tilde{B}^\mu A_\mu + \chi^2 A^\mu A_\mu). 
\end{multline}

\noindent Here, we have absorbed the renormalization of the electromagnetic gauge coupling $e^2 \rightarrow {e^2/(1 - \chi^2)}$. The mass term in~\equaref{Lint} coupling the electromagnetic and hidden sector vector potentials gives rise to a mixing angle between hidden sector interaction and mass eigenstates. This provides the mechanism for paraphoton-photon oscillations~\cite{Okun} analogous to that of neutrino flavors.

The coupling of the paraphoton to the photon modifies the electromagnetic potential with an additional Yukawa term~\cite{Bartlett}. Constraints from the non-observation of deviations from Coulombs law~\cite{Williams} and the black-body spectrum of the cosmic microwave background~\cite{CMB} exclude couplings $\chi>10^{-7}$ in the paraphoton mass range $10^{-9}<\Mpara<10^{-4}~\rm{eV}$.
 
Microwave cavities can be used as sensitive searches for paraphotons via the resonant enhancement of photons mixing into paraphotons in an emitter cavity and the observation of transmitted power in a detector cavity resulting from the paraphotons oscillating back into photons~\cite{JackelRing}. Such cavities typically cover the paraphoton mass range $10^{-6}<\Mpara<10^{-3}~\rm{eV}$. The power observed in the detector cavity $\Pdet$, for an emitter cavity supplied with power $\Pem$, is given by~\cite{JackelRing}

\begin{equation}
\label{eq:pow}
\Pdet=\chi^4Q_{\rm det}Q_{\rm em}G^2\bigg(\frac{\Mpara}{\omega}\bigg)^8\Pem,
\end{equation} 

\noindent where the emitter cavity is driven with frequency $\nu=\omega/2\pi$, and $Q_{\rm det},~Q_{\rm em}$ are the quality factors of the detector and emitter cavities respectively. These factors account for the resonant enhancement of photon paraphoton mixing probabilities. They are typically of order $10^4$ for cavities used by ADMX. The dimensionless coefficient $G$ accounts for the interference of photons in the detector cavity as paraphotons mix back in different locations with different relative phases and is given by 

\begin{equation}
\label{eq:G}
G=\omega^2 \int_{V_{\rm em}}\int_{V_{\rm det}}{e^{ik|\bold{x-y}|}\over {4\pi|\bold{x-y}|}}\vec{A}_{\rm em}(\bold{y})\cdot\vec{A}_{\rm det}(\bold{x})d^3\bold{y}d^3\bold{x}.
\end{equation} 

\begin{figure}[h]
  \begin{center}
    \begin{tabular}{lr}
      \epsfig{file=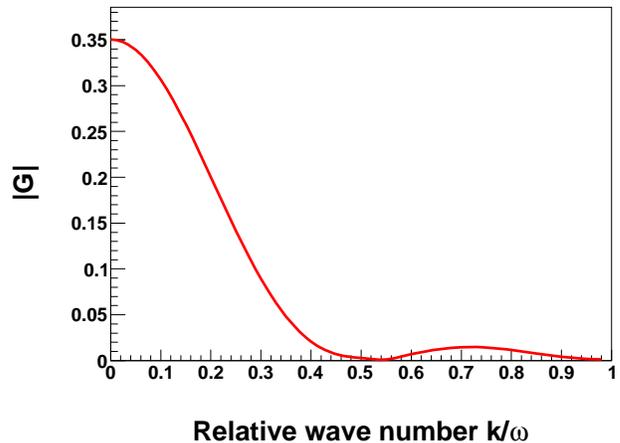,width=9cm}
    \end{tabular}
    \caption{\label{fig:G} Magnitude of the dimensionless coefficient $G$ given in~\equaref{G} as a function of relative paraphoton wave number. The curve is specific to this experiment~(\figref{Exp}).}
  \end{center}
\end{figure}

\begin{figure}[h]
  \begin{center}
    \begin{tabular}{lr}
      \epsfig{file=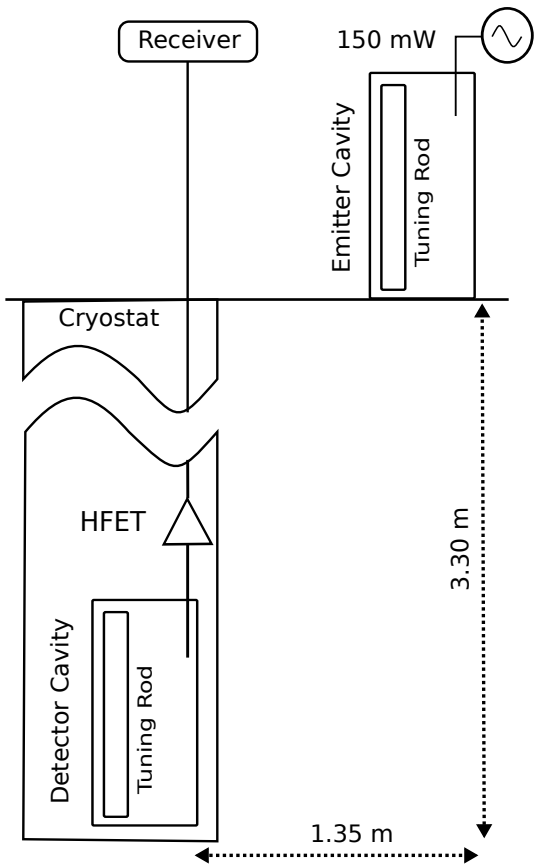,width=7cm}
    \end{tabular}
    \caption{\label{fig:Exp} Diagram of the ADMX paraphoton search. Photons mix into paraphotons in the emitter cavity and back into photons in the detector cavity. The power deposited by photons in the detector cavity is read out by an antenna and amplified within the cryostat.}
  \end{center}
\end{figure}

\noindent Here, $k$ is the wave number of the paraphoton, $k^2 = \omega^2 - \Mpara^2$, and the vector potentials of the EM normal modes, $\vec{A}_{\rm em},~\vec{A}_{\rm det}$, obey the normalization condition

\begin{equation}
\label{eq:A_norm}
\int_{V}|\vec{A}|^2(\bold{x})d^3\bold{x} = 1, 
\end{equation}

\noindent where we use the Coulomb gauge $\vec{\nabla}\cdot\vec{A}=0$. The magnitude of $G$ is shown in~\figref{G} as a function of the relative paraphoton wave number, $k/\omega$, for the arrangement of cavities in this experiment, shown in~\figref{Exp}. 

ADMX employs a tunable microwave cavity immersed in a 7.5 T magnetic field. It was originally built to detect the conversion of dark matter axions into photons. The ADMX detector is described in detail in Ref.~\cite{NIM}, and recent results of the axion search are given in Ref.~\cite{ADMX_prl}.

Microwave power from paraphoton mixing can be detected in an identical manner to that from axion conversions. Electromagnetic power in the ADMX cavity excites the \TMmode~normal mode which is read out by an antenna, amplified, mixed, and Fourier transformed with a frequency resolution of 125 Hz. The sensitivity of power measurements is limited by black body noise from the cavity and the intrinsic noise of the amplifier. It is thus important to operate at as low a temperature as possible. The present experiment maintains the cavity at a physical temperature of 2 K with pumped helium while the High electron mobility Field Effect Transistor (HFET) amplifiers~\cite{HEMT} have a noise temperature of 4 K. This results in a total noise temperature of 6 K. 

A significantly lower cavity temperature would reduce the cavity black body noise, but not the HFET noise. The use of Superconducting Quantum Interference Device (SQUID) amplifiers, however, with a noise temperature approaching the quantum limit at low temperatures~\cite{SQUID}, reduces the system noise temperature substantially, for example, to a projected 200 mK or less for an operating temperature of 100 mK. The SQUID amplifiers used during normal ADMX operations were not operational for this experiment. 

The signal-to-noise ratio for power detected in the ADMX cavity is set by the radiometer equation~\cite{Rad}

\begin{equation}
\label{eq:SNR}
{S\over N}={\Pdet\over k_BT}\sqrt{t\over b},
\end{equation} 

\noindent where $T$ is the total noise temperature, $t$ is the time over which the power is measured and $b$ is the frequency bandwidth resolution. A paraphoton signal would be observed as a peak above the noise spectrum in the detector cavity at the frequency of the driven emitter cavity. We find that the noise temperature of the system dominates any systematic effects of the receiver chain for integration times less than 30 days~\cite{ADMX_prdrc}. 

The paraphoton search procedure is to operate the ADMX detector cavity in the \TMmode~mode, tuned via the placement of two 0.0508 m diameter rods, to a resonant frequency of 722.725 MHz. A second cavity is supplied with 150 mW of power in the same mode and frequency. Both the cavities and tuning rods are copper plated stainless steel cylinders. The cavities each measure 0.40 m in diameter and 0.926 m in height and are separated 3.30 m vertically and 1.35 m horizontally due to the presence of the detector cryostat. The quality factors of the cavities, $Q_{\rm det},Q_{\rm em}$ are $7\times 10^4$ and $1.4\times 10^4$ respectively at 722 MHz. A schematic of the ADMX paraphoton search is shown in~\figref{Exp}. 

\begin{figure}[h]
  \begin{center}
    \begin{tabular}{lr}
      \epsfig{file=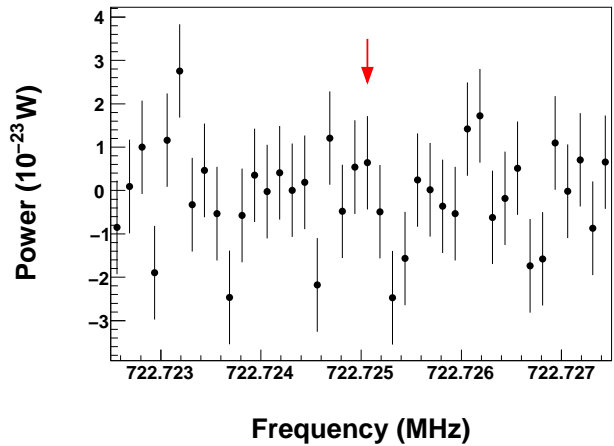,width=9cm}
    \end{tabular}
    \caption{\label{fig:power} Excess power measured in the ADMX cavity as a function of frequency in MHz. Bins are 125 Hz wide. The emitter cavity is driven at 722.725 MHz and paraphoton mixing would be detected as a peak above the noise power in that bin (arrow).}
  \end{center}
\end{figure}
       
\begin{figure}[h]
  \begin{center}
    \begin{tabular}{lr}
      \epsfig{file=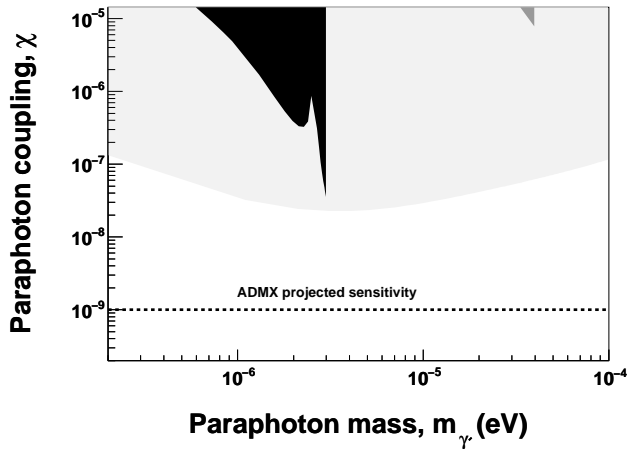,width=9cm}
    \end{tabular}
    \caption{\label{fig:limit} Limits on the kinetic coupling of the paraphoton as a function of mass at the $95\%$ confidence level. The ADMX limit is shown as dark shading. An earlier microwave cavity limit~\cite{Tobar} is shown as medium shading. The limit from Coulomb's law deviations~\cite{Bartlett,Williams} is shown as light shading.}
  \end{center}
\end{figure}

We average the power in a 125 Hz wide bin around 722.725 MHz for a period of $7.2\times10^3$~s. We limit the excess power to $\Pdet=0.69\pm1.08\times10^{-23}~{\rm W}$~(\figref{power}). Since this is consistent with no excess power we set a limit on the paraphoton coupling $\chi$ as function of mass, $\Mpara$ using~\equaref{pow} and~\figref{G}. We exclude paraphotons over a broad mass range. For masses near $3~\mueV$, we exclude kinetic couplings $\chi > 3.48\times 10^{-8}$ at the $95\%$ confidence level as shown in~\figref{limit}. We exclude couplings to lower paraphoton masses with reduced sensitivity to the coupling.

Feed-through of microwave power from the emitter to the receiver antennas is a potentially serious systematic effect that dramatically limited the sensitivity of microwave cavities to paraphoton mixing in a previous experiment~\cite{Tobar}. However; ADMX has been optimized to eliminate external microwave signals which could limit the sensitivity of the axion search. The ADMX cryostat functions as an excellent Faraday cage where all signal and sensor lines have been tested for leakage with an external microwave source. We have terminated all unnecessary lines and supplied the remainder with additional shielding. The HFET amplifiers supply 30 dB of amplification and are well shielded within the cyrostat. This greatly suppresses the effect of any feed-through from outside the cryostat. Consequently, the ADMX paraphoton search was not limited by the feed-through of microwave power. 

The ADMX receiver will be upgraded to improve the experiment's sensitivity. This upgrade will employ SQUID amplifiers with noise temperatures close to the quantum limit and a dilution refrigerator cooled to 100 mK to reduce the total noise temperature to 200 mK. Improved frequency stability will allow bandwidth resolution of better than 0.02 Hz. The reinstallation of ADMX at the University of Washington will also enable the emitter and detector cavities to be positioned within 1.5 m so that $|G|\approx 1$. The power driving the emitter cavity was supplied by the source used for nominal ADMX operations; power levels in excess of 1 W will be available in the receiver upgrade. We estimate this upgrade will provide sensitivity to paraphotons with a kinetic coupling $\chi< 10^{-9}$~(\figref{limit}).  

In summary, we have used the ADMX detector in a search for hidden sector physics. We exclude paraphotons over a broad mass range. For paraphoton masses near $3~\mueV$ we exclude kinetic couplings $\chi > 3.48\times 10^{-8}$ at the $95\%$ confidence level. Lower paraphoton masses are excluded with reduced sensitivity to the kinetic coupling. Upgrades to ADMX will permit an improvement in this limit by more than an order of magnitude. This measurement is a significant improvement in microwave cavity limits and is comparable to the sensitivity of the limit set in Refs.~\cite{Bartlett,Williams} by tests of Coulomb's law. Finally, this experiment, together with the axion~\cite{ADMX_prl} and chameleon~\cite{ADMX_chameleon} searches performed by ADMX, demonstrates not only the extraordinary sensitivity of microwave cavity experiments but also the breadth of physics such experiments can expolore.   

The ADMX collaboration gratefully acknowledges support by the U.S. Department of Energy, Office of High Energy Physics under contract numbers DE-FG02-96ER40956 (University of Washington), DE-AC52-07NA27344 (Lawrence Livermore National Laboratory), and DE-FG02-97ER41029 (University of Florida). Additional support was provided by Lawrence Livermore National Laboratory under the LDRD program. Development of the SQUID amplifier (JC) was supported by the Director, Office of Science, Office of Basic Energy Sciences, Materials Sciences and Engineering Division, of the U.S. Department of Energy under Contract No. DE-AC02-05CH11231.

\bibliographystyle{h-physrev}

\begin{thebibliography}{99}

\bibitem{HamedWeiner} N. Arkani-Hamed, D.P. Finkbeiner, T.R. Slatyer and N. Weiner, \jprd{79}, 015014 (2009). 

\bibitem{Holdom} B. Holdom, \plb{166}, 196 (1986).

\bibitem{PAMELA} O. Adriani \ea, \nature{458}, 607 (2009). 

\bibitem{hCMB} J. Jaeckel and J. Redondo and A. Ringwald, \jprl{101}, 131803 (2008).

\bibitem{JackelRing} J. Jaeckel and A. Ringwald, \plb{659}, 509 (2008).

\bibitem{Okun} L.B. Okun, \spjetp{56}, 502 (1982).

\bibitem{Bartlett} D.F. Bartlett and S. L\"{o}gl, \jprl{61}, 2285 (1988).

\bibitem{Williams} E. R. Williams, J. E. Faller and H. A. Hill, \jprl{26}, 721 (1971). 

\bibitem{CMB} A. Mirizzi, J. Redondo and G. Sigl, JCAP {\bf 0903}, 026 (2009).

\bibitem{NIM} H. Peng \ea, \nima{444}, 569 (2000).

\bibitem{ADMX_prl} S. J. Asztalos \ea, \jprl{104}, 041301 (2010). 

\bibitem{HEMT} E. Daw and R. Bradley, \jap{82}, 1925 (1997). 

\bibitem{SQUID} M. M\"{u}ck, J.B. Kycia and J. Clarke, \aplet{78}, 967 (2001). 

\bibitem{Rad} R. H. Dicke, \rsi{17}, 268 (1946). 

\bibitem{ADMX_prdrc} S. J. Asztalos \ea, \jprd{69}, 011101(R) (2004). 

\bibitem{Tobar} R. Povey, J. Harnett, and M. Tobar, \verb!{arXiv:hep-ex/1003.0964}!.

\bibitem{ADMX_chameleon} G. Rybka \ea, \jprl{105}, 051801 (2010).  


\end{thebibliography}

\end{document}